\documentclass[dvips,twocolumn,secnumarabic,amssymb,nobibnotes,aps,prl]{revtex4-1}    % sub
\usepackage{amsmath}    % need for subequations
\usepackage{graphicx}   % need for figures
\usepackage{verbatim}   % useful for program listings
\usepackage{xspace}

\usepackage[caption=false]{subfig}
\DeclareCaptionListOfFormat{myparens}{#2}
\newcommand{\Subref}[1]{\protect\subref{#1}}

\input{aas_macros.sty}

\usepackage[dvips,dvipsnames,usenames]{color}
\usepackage{varioref}
\usepackage{colortbl}
\usepackage[dvips, breaklinks={true}, bookmarks, colorlinks={true}, pdfpagemode={NON}, pdffitwindow=true, pdfstartview={FitB}, linkcolor={BlueViolet}, menucolor={BlueViolet}, citecolor={Mahogany}, filecolor={OliveGreen}, pagecolor={Mahogany}, urlcolor={BlueViolet}, pdftitle={Gravitational redshift of galaxies in clusters from SDSS and BOSS}, pdfauthor={Iftach Sadeh}, pdfsubject={Gravitational redshift of galaxies in clusters}, pdfkeywords={gravitational redshift, galaxy clusters, SDSS, BOSS}]{hyperref}

\let\orgautoref\autoref

\renewcommand{\autoref}
        {\def\equationautorefname{Eq.}%
         \def\figureautorefname{Fig.}%
         \def\subfigureautorefname{Fig.}%
         \def\chapterautorefname{Ch.}%
         \def\sectionautorefname{Sect.}%
         \def\subsectionautorefname{Sect.}%
         \def\subsubsectionautorefname{Sect.}%
         \def\Itemautorefname{item}%
         \def\tableautorefname{Table}%
         \orgautoref}

\newcommand{\eg}        {e.g.\/,\xspace}
\newcommand{\etal}      {\textit{et~al.\ \xspace}}

\newcommand{\powA}[2]{\ensuremath{{{#1}\cdot10^{{#2}}}}\xspace}
\newcommand{\powB}[1]{\ensuremath{{10^{{#1}}}}\xspace}

\newcommand{\mrm}[1]{\mathrm{#1}}

\newcommand{\mSun}{\ensuremath{M_{\odot}}\xspace}
\newcommand{\vGc}{\ensuremath{v_{\mrm{gc}}}\xspace}
\newcommand{\dV}{\ensuremath{\Delta v_{\mrm{gc}}}\xspace}
\newcommand{\nGal}{\ensuremath{n_{\mrm{gal}}}\xspace}

\newcommand{\nClst}{\ensuremath{n_{\mrm{clst}}}\xspace}
\newcommand{\mrGc}{\ensuremath{\delta m^{\mrm{gc}}_{r}}\xspace}
\newcommand{\rGc}{\ensuremath{r_{\mrm{gc}}}\xspace}
\newcommand{\rV}{\ensuremath{r_{200}}\xspace}

\newcommand{\mNest}{\texttt{MultiNest}\xspace}

\newcommand{\lcdm}{\ensuremath{\Lambda}CDM\xspace}
\newcommand{\kms}{\ensuremath{\mrm{km/s}}\xspace}

\begin{document}
% --------------------------------------------------------------------------------------------

\title{Gravitational Redshift of Galaxies in Clusters from the Sloan Digital\\Sky Survey and the Baryon Oscillation Spectroscopic Survey}

\author{Iftach Sadeh}%
\email[Iftach Sadeh,~]{i.sadeh@ucl.ac.uk}
\author{Low Lerh Feng}%
\email[Low Lerh Feng,~]{lerh.low.13@ucl.ac.uk}
\author{Ofer Lahav}%
\email[Ofer Lahav,~]{o.lahav@ucl.ac.uk}
\affiliation{Astrophysics Group, Department of Physics and Astronomy, University College London, Gower Street, London WC1E 6BT, United Kingdom}
\date{October 20, 2014}

\begin{abstract}
The gravitational redshift effect allows one to directly probe the gravitational
potential in clusters of galaxies.
Following up on Wojtak \etal [Nature (London) 477, 567 (2011)], we present
a new measurement.
We take advantage of new data from the tenth data release of
the Sloan Digital Sky Survey and the Baryon Oscillation Spectroscopic Survey.
We compare the spectroscopic redshift of the
brightest cluster galaxies (BCGs) with that of galaxies
at the outskirts of clusters, using a sample with an average cluster mass of ${\powB{14}\mSun}$.
We find that these galaxies have an average relative redshift of~${-11~\kms}$
compared with that of BCGs, with a standard deviation of~$+7$ and $-5~\kms$.
Our measurement is consistent with that of Wojtak \etal However,
our derived standard deviation is larger, 
as we take into account various systematic effects, beyond the size of the dataset.
The result is in good agreement with the predictions from general relativity.
\end{abstract}

% The gravitational redshift effect allows one to directly probe the gravitational potential in clusters of galaxies. Following up on Wojtak et al. [Nature (London) 477, 567 (2011)], we present a new measurement. We take advantage of new data from the tenth data release of the Sloan Digital Sky Survey and the Baryon Oscillation Spectroscopic Survey. We compare the spectroscopic redshift of the brightest cluster galaxies (BCGs) with that of galaxies at the outskirts of clusters, using a sample with an average cluster mass of $10^{14} M_{\odot}$. We find that these galaxies have an average relative redshift of -11 km/s compared with that of BCGs, with a standard deviation of +7 and -5 km/s. Our measurement is consistent with that of Wojtak et al. However, our derived standard deviation is larger, as we take into account various systematic effects, beyond the size of the dataset. The result is in good agreement with the predictions from general relativity.

\maketitle

% --------------------------------------------------------------------------------------------
\section{Introduction}
% --------------------------------------------------------------------------------------------
%
The gravitational redshift (GRS) effect in clusters of galaxies
is a feature in any metric theory of gravity. It is
caused by the spatial variation of the gravitational potential;
light traveling from deeper in the potential of a cluster
is expected to be redshifted, compared 
to light originating from the outskirts of the
cluster~\cite{2004ApJ...607..164K,*1983A&A...118...85N,*1995A&A...301....6C}.
The GRS effect has the potential to constrain 
theories in which there are long-range non-gravitational forces acting on dark
matter, modifying gravity on cluster
scales~\cite{1993ApJ...403L...5G,*2004PhRvD..70l3511G,*2010PhRvD..81f3521K,*2007PhRvL..98q1302F}.
The effect was first measured by Wojtak, Hansen \& Hjorth (WHH)~\cite{2011Natur.477..567W},
a study which was subsequently repeated
(with minor modifications) by Dominguez-Romero \mbox{\etal \cite{MNL2:MNL21326}}.

WHH used 125k~spectroscopic redshifts,
taken from the seventh data release~\cite{Abazajian:2008wr}
of the Sloan Digital Sky Survey (SDSS)~\cite{2011AJ....142...72E},
matched to 7.8k~clusters from the GMBCG cluster catalog~\cite{2010ApJS..191..254H}.
They used the brightest cluster galaxies (BCGs) as a proxy
for the centers of clusters. They assumed that, in general, BCGs
have relatively small velocity dispersions
compared to other bound galaxies,
and reside close to the bottom of the gravitational potential.
WHH divided their galaxy sample into four bins, based on the transverse distance between cluster-galaxies
and respective BCGs, \rGc, extending up to~${6~\mrm{Mpc}}$. In each bin, they calculated the
line-of-sight velocity of galaxies in the rest-frame of the BCG,
\begin{equation}
  \vGc = c \; \frac{z_{\mrm{gal}} - z_{\mrm{BCG}} }{1 + z_{\mrm{BCG}}} \;,
\label{eq_vGcDef}
\end{equation}
where $z_{\mrm{BCG}}$ and $z_{\mrm{gal}}$ respectively stand for the redshift
of BCGs and of associated galaxies, and $c$ is the speed of light.
The stacked \vGc-distributions of galaxies from the entire sample of clusters
were fitted with a phenomenological model, using a Markov chain Monte Carlo (MCMC) program.
The derived mean of the distributions was interpreted as the GRS signal.
On average, cluster-galaxies were found to have a
redshift difference relative to corresponding BCGs, 
amounting to a velocity difference, $\dV\approx-7~\kms$.

WHH calculated the prediction for the signal from general relativity (GR), as well as from
modified theories of gravity~\cite{2004PhRvD..70d3528C,*1983ApJ...270..365M,*2004PhRvD..70h3509B}.
They first derived the GRS profile of a single cluster in the weak field limit,
\begin{equation}
  \Delta_{1}(\rGc) = \frac{2}{c \; \Sigma(\rGc)}\int\limits_{\rGc}^{\infty}
               \left[\Phi(r) - \Phi(0) \right]
               \frac{ \rho(r) \; r }{ \sqrt{\smash[b]{r^2 - \rGc^2}} } \mrm{d}r \;,
\label{eq_whhGR}
\end{equation}
where $\Phi$ is the gravitational potential, and $\rho$ and $\Sigma$ are respectively the
three-dimensional and surface-density profiles of galaxies. They then convolved $\Delta_{1}$ with
the distribution of cluster masses in their sample, estimated from
the observed velocity dispersion profile, using stacked NFW models~\cite{1997ApJ...490..493N}.

Subsequent works, notably those of Zhao~\etal\cite{2013PhRvD..88d3013Z}  and of Kaiser~\cite{2013MNRAS.435.1278K},
modified the theoretical prediction; these took into account effects such as
the so-called transverse Doppler shift and
surface brightness modulation. The added
corrections were found to be of the same order of magnitude as the GRS
signal, some inducing redshifts and some blueshifts. Summed together, the prediction
of Kaiser is of a relatively flat dependence of \dV on \rGc,
with a mean value of~$-9$ (GR only)
or~$-12~\kms$ (GR and kinematic effects).

The purpose of this study is to revise the measurement of WHH.
In the next sections we describe the analysis in detail, following up with our results
and conclusions.

% --------------------------------------------------------------------------------------------
\section{Methodology}
% --------------------------------------------------------------------------------------------
%
% --------------------------------------------------------------------------------------------
\subsection{Dataset}
% --------------------------------------------------------------------------------------------
%

We used spectroscopic redshifts derived from the tenth data release (DR10)~\cite{dr10.1307.7735} of the SDSS,
including measurements taken with the Baryon Oscillation Spectroscopic Survey (BOSS)~\cite{boss.1208.0022},
occupying the redshift range, $0.05$~to~$0.6$.
We associated the DR10 data with galaxy clusters, using
the catalog of Wen, Han \& Liu (WHL)~\cite{2012ApJS..199...34W}.
The WHL sample includes~${\sim130\mrm{k}}$
clusters, detected using a friends-of-friends algorithm, based on photometric data.
The virial radius of a cluster is commonly approximated by \rV,
the radius within which the mean density of a cluster is
200~times that of the critical density of the universe~\cite{peebles1993principles,*peacock1999cosmological}.
The latter is additionally
used to define  $m_{200}$, the cluster mass within \rV.
The WHL catalog is nearly complete for clusters with masses, ${m_{200}>\powA{2}{14}\mSun}$,
and redshifts, ${z<0.5}$, and is~${\sim75\%}$ complete for ${m_{200}>\powA{0.6}{14}\mSun}$
and ${z<0.42}$.
Cluster mass is estimated using a scaling relation between mass and
optical richness.
The latter was estimated by WHL using x-ray or weak-lensing
methods, and is given in ${\mrm{Eq.\;2}}$ of~\cite{2012ApJS..199...34W}.
The derived average cluster mass in our selected cluster sample
is ${m_{200} = \powA{1.3}{14}\mSun}$.
This is commensurate with the mean value of cluster masses in the WHH
dataset, ${m_{200} = \powA{1.6}{14}\mSun}$, allowing us to directly
compare the results of their measurement with our own.

In the initial stage of the analysis, the
spectroscopic redshifts were subjected to various quality cuts,
ensuring \eg that the uncertainty on the redshift is below~\powB{-4}, and that the
confidence in the likelihood-fit of the redshift is high.
We then matched galaxy spectra to BCG positions,
keeping only those clusters for which the BCG had a corresponding spectrum.
Additionally, each cluster had to contain
at least one galaxy within transverse distance, ${\rGc<6~\mrm{Mpc}}$,
and velocity, ${\left| \vGc \right| < 4,000~\kms}$.
Conversion from angular to physical distances was performed using a flat \lcdm cosmology, with
${\Omega_{\mrm{m}} = 0.307}$ and the Hubble
constant, ${\mrm{H}_{0} = 67.8~\mrm{km}\;\mrm{s}^{-1}\;\mrm{Mpc}^{-1}}$~\cite{2014A&A...571A..16P}.

The initial selection left us
with~${31\mrm{k}}$ clusters and~${426\mrm{k}}$ associated galaxies.
% which we denote as our baseline sample.
Following the selection procedure discussed in the next section, we were left with
${60\mrm{k}}$ galaxies and ${12\mrm{k}}$ clusters, having ${\rGc \lesssim 3~\mrm{Mpc}}$. An
additional ${25\mrm{k}}$ galaxies and ${5\mrm{k}}$ clusters were
used for systematic checks.

% --------------------------------------------------------------------------------------------
\subsection{Fitting procedure}
% --------------------------------------------------------------------------------------------
%
WHH employed a MCMC program to fit the velocity distribution to the phenomenological model,
\begin{equation}
  f(\vGc) = p_{\mrm{cl}} \cdot f_{\mrm{Gauss}}(\vGc) + \left( 1 - p_{\mrm{cl}} \right)\cdot f_{\mrm{Lin}}(\vGc) \;,
\label{eq_vGcFitModel}
\end{equation}
where $f_{\mrm{Gauss}}$ is a
convolution of two Gaussian distributions, having a common mean value, \dV,
and $f_{\mrm{Lin}}$ is a linear function.
The quasi-Gaussian
contribution represents galaxies bound to clusters.
It accounts for the 
intrinsic non-Gaussianity of velocity distributions of individual clusters,
and for the variation in cluster masses in the sample.
The linear part of the model represents a uniform background of interlopers (line-of-sight
galaxies which are not gravitationally bound to the cluster).
The fraction of bound galaxies,
$p_{\mrm{cl}}$, is a free parameter of the MCMC program, which is marginalized over,
as are the two coefficients of $f_{\mrm{Lin}}$,
the width of the two Gaussian functions, and the relative normalization of the two.

Scaling the separation between galaxies and associated BCGs by \rV
takes advantage of the self-similarity of clusters;
we therefore used \rGc-bins defined in units of \rV.
We fitted the \vGc-distribution with \autoref{eq_vGcFitModel} in each bin using \mNest,
a Bayesian inference tool employing importance nested
sampling~\cite{2009MNRAS.398.1601F, *2008MNRAS.384..449F, *2013arXiv1306.2144F}.
The fits for the various \rGc-bins
were found to be compatible with the data, scoring better than~$99\%$ in K-S tests.

The observed velocity dispersions were of the order of
several hundred~\kms, more than 50~times larger than the 
GRS signal, \dV. As the signal was difficult to confirm visually,
we also computed the ratio between the integral of the negative and of
the positive parts of the \vGc-distribution. The latter is a model-independent
measure of the magnitude of the signal. It was shown to correlate well with
the derived value of \dV, validating that the \mNest fitting procedure
is not biased.
In addition, we wrote a simple Metropolis-Hastings MCMC program
and cross-checked the fit-results.

% --------------------------------------------------------------------------------------------
\subsection{Sample composition and systematic tests}
% --------------------------------------------------------------------------------------------
%

%  - - - - - - - - - - - - - - - - - - - - - - - - - - - - - - - - - - - - - - - - - - - - - -
Our baseline dataset may be utilized in various ways to perform the measurement. One of the
main sources of ambiguity is that we are interested in galaxies which are
several~Mpc away from the corresponding BCGs.
On average, the distance between close pairs of clusters in our dataset
corresponds to~${2.3  \rV}$, where for
the bulk of the cluster sample, ${0.8<\rV<1.2~\mrm{Mpc}}$.
Many galaxies are therefore likely to be associated with multiple clusters, depending on
their extent and separation.

We nominally define a pair of overlapping
clusters as having a transverse separation, ${r_{\mrm{cc}}<4\rV}$,
and a velocity difference, ${\left|v_{\mrm{cc}} \right| < 4,000~\kms}$.
We tested several galaxy selection schemes, with different restrictions
on overlapping configurations.
One possible selection procedure is to exclude all overlapping cluster pairs from
the analysis. Another option is to exclude all but one member from any
configuration of overlapping clusters.
Alternatively, we may choose not to take
into account cluster overlaps at all.
We then accept only those galaxies that have only one cluster association, effectively
performing exclusive selection on galaxies instead of on clusters.

%  - - - - - - - - - - - - - - - - - - - - - - - - - - - - - - - - - - - - - - - - - - - - - -
In order to check the dependence of the signal on the composition of our dataset, we
ran the analysis on subsets of the data. 
Of these, tests involving
BOSS-BCGs associated with SDSS galaxies
revealed a systematic positive bias of a few~\kms.
So as to understand this effect, we 
define the quantity, ${\mrGc = m_{r}^{\mrm{g}} - m_{r}^{\mrm{c}}}$,
where $m_{r}^{\mrm{c}}$ and $m_{r}^{\mrm{g}}$ respectively stand for
the $r$-band magnitude of a BCG,
and that of the brightest matched galaxy within one \rV of the BCG.
Positive \mrGc values correspond to BCGs which are indeed found to be
the brightest source within the area of a cluster.
We observed on average, ${\mrGc = -0.3}$
for configurations in which the two surveys were mixed. 
The implication of this is that for this sub-sample, it is likely that BCGs
were misidentified in the cluster catalog.
As a result, selected BCGs were less likely to represent the bottom of the
gravitational potential well of clusters, effectively suppressing the GRS signal.
One should also keep in mind that the difference between SDSS and BOSS
redshifts is almost an order of magnitude smaller than the uncertainties on the
redshifts. It is therefore possible that the bias
originates \eg from changes made in the template-fitting procedure between data-releases.

%  - - - - - - - - - - - - - - - - - - - - - - - - - - - - - - - - - - - - - - - - - - - - - -
Another important systematic is the treatment of clusters with high
galaxies-multiplicities,
which we denote by \nGal. These configurations are subject to two types of bias.
The first is due to the fact that \vGc depends on the redshifts of both a
galaxy and the corresponding BCG; consequently, an error
in the redshift of a given BCG affects all matched galaxy-velocities in a correlated way.
The second effect that we observed, was that the value of \mrGc,
while generally positive, tends to decrease as \nGal increases.
We, therefore, concluded that clusters become more susceptible to misidentification of the BCG
with growing multiplicities.
In order to mitigate these effects, we down-weighted the contribution
of clusters with high multiplicities in the \vGc-distribution. We found that
this change mainly affected the signal for low values of \rGc.

%  - - - - - - - - - - - - - - - - - - - - - - - - - - - - - - - - - - - - - - - - - - - - - -
An additional possible source of bias is the uncertainty associated
with individual spectroscopic redshifts. We checked that there
was no correlation between these, and the corresponding values of \vGc.

% --------------------------------------------------------------------------------------------
\section{Results}
% --------------------------------------------------------------------------------------------
%
%
\begin{figure*}[htb]
\begin{center}
  \begin{minipage}[c]{0.385\textwidth}
    \subfloat[]{\label{nSpecClstFIG}\includegraphics[trim=9.5mm 80mm 25mm 0mm,clip,width=1.\textwidth]{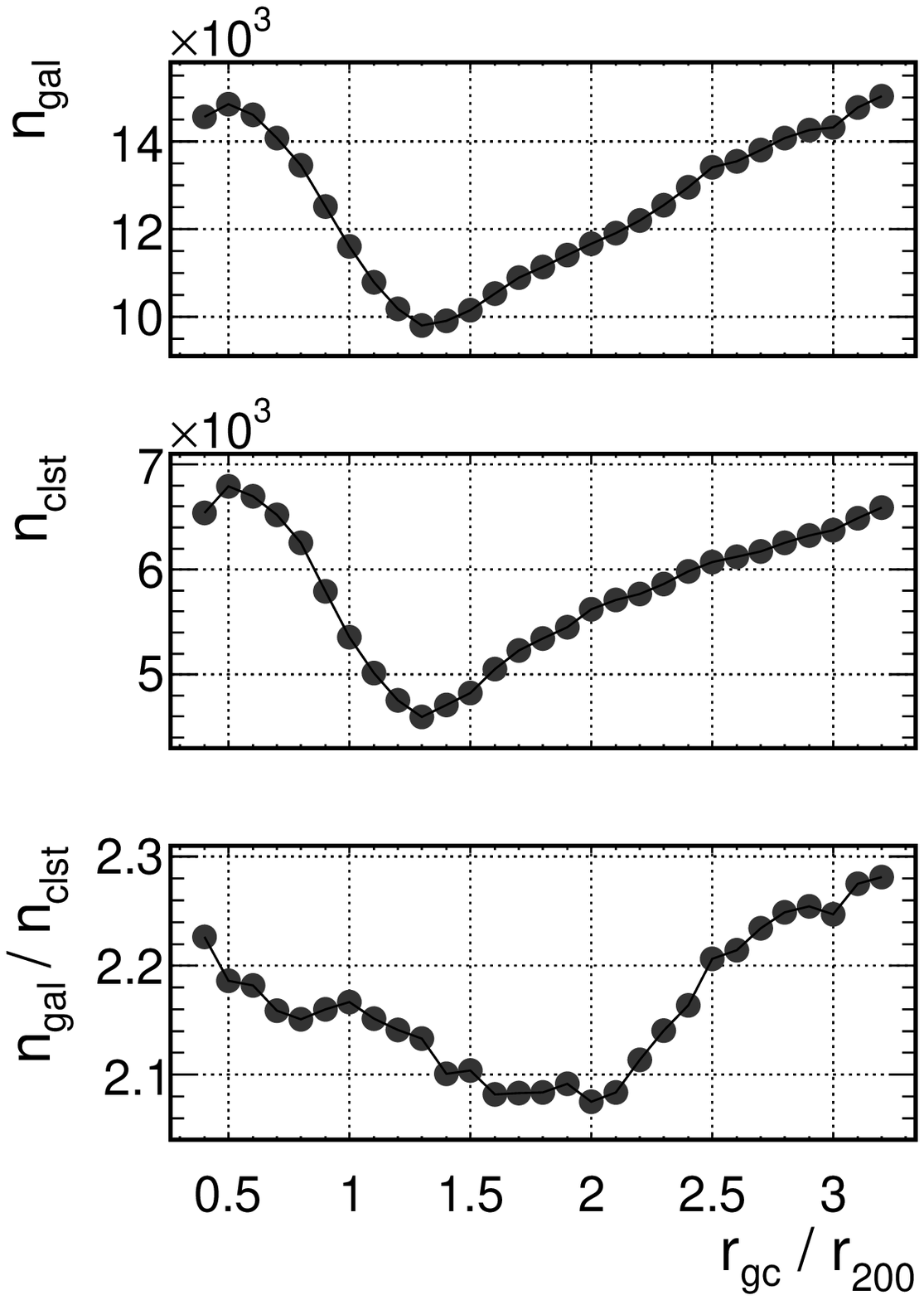}}
  \end{minipage}\hfill
  \begin{minipage}[c]{0.58\textwidth}
    \subfloat[]{\label{deltavGcFIG}\includegraphics[trim=3mm 0mm 20mm 0mm,clip,width=1.\textwidth]{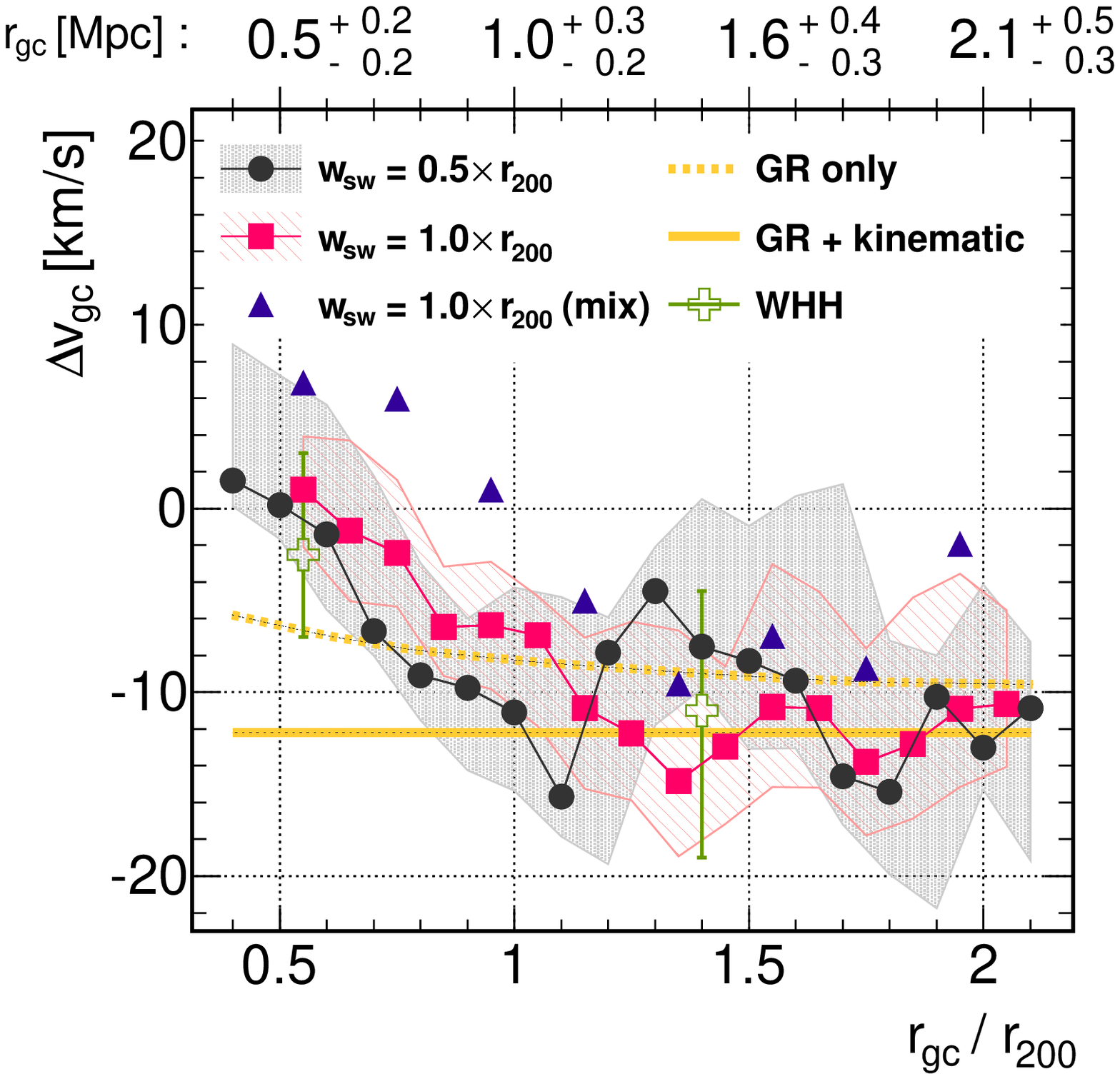}}
  \end{minipage}\hfill
  \caption{
    \Subref{nSpecClstFIG}:~Dependence of the number of galaxies, \nGal,
      of the number of associated clusters, \nClst, and of the
      ratio, ${\nGal/\nClst}$, on the separation between BCGs
      and associated galaxies, \rGc. Bins of \rGc are defined by a sliding window with
      a width of ${0.5  \rV}$, where each data-point is placed in the center-position of
      the corresponding bin. \\
    \Subref{deltavGcFIG}:~ Dependence of the signal of the GRS, \dV,
    on \rGc, where the width of the sliding window is denoted by $w_{\mrm{sw}}$.
    The shaded areas around the two
    nominal results (circles and squares) correspond
    to the variations in the signal due to the systematic tests described
    in the text, combined with the uncertainty on the model-fit.
    On average, ${\dV = {-11^{+7}_{-5}~\kms}}$ for ${1 < \rGc/\rV < 2.5}$.
    The third dataset (triangles)
    includes configurations in which SDSS and BOSS redshifts are mixed together.
    The bold lines represent the GR predictions of Kaiser~\cite{2013MNRAS.435.1278K},
    with and without his added kinematic effects, as indicated; finally, the crosses
    represent the measurement of WHH.
    The top axis specifies the median
    value and the width of the distribution of \rGc (in Mpc) for four bins of
    width ${0.5  \rV}$, centered at
    ${\left( 0.5,\;1,\;1.5\;\mrm{and}\;2 \right) \rV}$.
  } \label{nSpecClst_deltavGcFIG}
\end{center}
\end{figure*} 

We estimated \dV using a sliding window 
for the transverse separation between galaxies and clusters. The sliding
window nominally had a width of~${0.5  \rV}$, and a step size of~${0.1  \rV}$.
For our primary selection, we elected to discard configurations in which SDSS and
BOSS spectra were mixed together.
This reduced the number of clusters and galaxies by~$16\%$ and~$11\%$, respectively.
We also excluded all overlapping cluster pairs,
further reducing these numbers by a respective~$34\%$ and~$46\%$.
The final dataset was composed
of ${12600}$ clusters and ${60626}$ matched galaxies. The
measurement was restricted to
transverse separation values below~${2.5  \rV}$.

The reason for this last condition may be inferred from \autoref{nSpecClstFIG}.
The figure shows the dependence on \rGc of the number of galaxies, \nGal,
of the number of associated clusters, \nClst, and of the
ratio, ${\nGal/\nClst}$. One may observe that for ${\rGc < 1.3  \rV}$,
the multiplicities of matched clusters and galaxies decrease; this is in accordance with the
expected trend for the surface density of galaxies in clusters
(see \eg figure~8 in~\cite{2005ApJ...633..122H}). However, for ${\rGc \gtrsim 2  \rV}$,
both the multiplicities and the galaxy-to-cluster ratio increase. This comes about
as galaxies at large \rGc have an increasingly higher probability
of being associated with another cluster. Such configurations therefore
tend to suppress the signal of the GRS, and should be rejected from the analysis.

The dependence of \dV on \rV is presented in \autoref{deltavGcFIG}.
We find that on average, ${\dV = {-11^{+7}_{-5}~\kms}}$ for ${1 < \rGc/\rV < 2.5}$,
with uncertainties given as 1~standard deviation of the average signal.
In physical scales, the measurement extends up to
galaxy-cluster transverse separations of~${\sim3~\mrm{Mpc}}$.

In addition to the primary result, we present a measurement with an increased
\rGc-bin width.
The two results are consistent within uncertainties.
We note that the set-up with the wider bins has a radial resolution
which is slightly too low to describe
the GRS effect at low values of \rGc. On the other hand, for high \rGc-values,
the increase in statistics in each bin seems to stabilize the result.
Finally, we also include a measurement of \dV, in which SDSS and BOSS
spectra are used congruently.
This change incurs a systematic shift of a few~\kms, as discussed above.

The uncertainty on \dV was derived from that on the \mNest fit,
combined with the variations incurred
due to the following systematic checks:
changing the minimal number of matched galaxies in a cluster
between~1 and~7;
not down-weighting clusters with high galaxy multiplicities;
using different overlap-removal methods, as described above;
changing the values of the cluster overlap parameters, using
${3 < r_{\mrm{cc}}/\rGc < 5}$ and ${3000 < v_{\mrm{cc}} < 6000~\kms}$;
down-weighting galaxies with high spectroscopic redshift uncertainties;
randomly excluding a fraction of galaxies or clusters of a given
data sample. % (\ie excluding $10\%$ of the SDSS-\main spectra).
Of these, the dominant systematic variation originated from changing $r_{\mrm{cc}}$, the threshold for
the transverse separation between clusters.

For comparison, \autoref{deltavGcFIG} also shows the results of WHH within the region of interest.
Our measurement is consistent with these, and has comparable uncertainty estimates.
However, we note that the significance of our result is smaller than
that of WHH, who quote a value, ${\dV = -7.7 \pm 3.0~\kms}$.
The reason for this is that
WHH computed the integrated signal for all clusters within ${\rGc < 6~\mrm{Mpc}}$. They estimated
the uncertainty from that of their MCMC model-fit, which was mainly determined by the size
of their data sample.
In the case of the current analysis, the uncertainty is driven by our 
systematic tests, rather than by the available number of clusters and galaxies.
Considering these, and the limited range of acceptance in \rGc, our
final relative uncertainty on \dV is higher.

Calculating the GR predication for \dV is beyond the scope of this study. However,
the range of cluster masses used in our analysis is comparable to that of the WHH sample.
We therefore refer to the corresponding estimate of Kaiser
of~$-9$ (GR only) or~${-12~\kms}$ (including kinematic effects)~\cite{2013MNRAS.435.1278K}.
Our results are in good agreement with this prediction for ${\rGc > \rV}$,
while at smaller values of \rGc, the profile of \dV is steeper in the data.
Additionally, we observe that
it is not possible to distinguish between the GR predictions with and without the
kinematic corrections.

% --------------------------------------------------------------------------------------------
\section{Summary}
% --------------------------------------------------------------------------------------------
%
The gravitational redshift effect allows one to directly probe the gravitational
potential in clusters of galaxies. As such, it provides a fundamental test of
GR.

Following up on the analysis of Wojtak, Hansen \& Hjorth,
we present a new measurement with a larger dataset.
We use spectroscopic redshifts
taken with the SDSS and BOSS, and match them to the BCGs of clusters
from the catalog of Wen, Han \& Liu.
The analysis is based on extracting the GRS signal from the distribution
of the velocities of galaxies in the rest frame of corresponding BCGs.
We focus on optimizing the selection procedure
of clusters and of galaxies, and take into account multiple
possible sources of systematic biases not considered by WHH.

We find an average redshift of~${-11~\kms}$
with a standard deviation of~$+7$ and $-5~\kms$ for ${1 < \rGc/\rV < 2.5}$. The result is
consistent with the measurement of WHH.
However, our overall systematic uncertainty
is relatively larger than that of WHH, mainly due to
overlapping cluster configurations; the
significance of detecting the GRS signal in the current analysis
is therefore reduced in comparison.
Our measurement is in good agreement with the GR predictions. 
Considering the current uncertainties, we can not distinguish between the
baseline GR effect and the recently proposed kinematic modifications.

With the advent of future spectroscopic surveys, such as
Euclid and DESI~\cite{2011arXiv1110.3193L,*desiRef},
we will have access to larger, more homogeneous datasets.
We expect that the new spectra will help to
reduce the systematic uncertainties on the measurement,
though dedicated target selection may be required.
Additionally, new data will facilitate novel techniques of
detecting the GRS signal, such as
the cross-correlation method suggested in~\cite{2013MNRAS.434.3008C}.

% --------------------------------------------------------------------------------------------
\section*{Acknowledgements}

% --------------------------------------------------------------------------------------------
%
We would like to thank 
Jacob Bekenstein,
Jens  Hjorth,
Pablo Jimeno,
Nick  Kaiser,
John  Peacock,
David Schlegel and
Radek Wojtak, 
for the useful discussions regarding the nature of 
spectroscopic redshifts, galaxy clusters and GRS.

O.L.\ acknowledges an Advanced European Research Council Grant, which
supports the postdoctoral fellowship of I.S.

This work uses publicly available data from the SDSS.
Funding for SDSS-III has been provided by the Alfred P.~Sloan Foundation,
the Participating Institutions, the National Science Foundation, and the U.S.\
Department of Energy Office of Science. The SDSS-III website
is \href{http://www.sdss3.org/}{http://www.sdss3.org/}.

\bibliographystyle{apsrev4-1} \bibliography{bib}

% --------------------------------------------------------------------------------------------
\end{document}